Original Paper

# Integration of Remote Patient Monitoring Systems into Physicians Work in Underserved Communities: Survey of Healthcare Provider Perspectives

Samuel Bonet Olivencia, M.S[1]; Karim Zahed, M.S.[1]; Farzan Sasangohar, Ph.D.[1,2]*Rotem Davir[3] and Arnold Vedlitz[3]

[1]Department of Industrial & Systems Engineering, Texas A&M University, College Station, TX 77843, USA; samuel089@tamu.edu; k.zahed@tamu.edu
[2]Center for Outcomes Research, Houston Methodist, Houston, TX 77030, USA
[3]Institute for Science, Technology and Public Policy (ISTPP), The Bush School of Government and Public Policy, Texas A&M University, College Station, TX 77843, USA; rdvir@tamu.edu; avedlitz@tamu.edu
*Correspondence: sasangohar@tamu.edu

**Abstract**

Remote patient monitoring (RPM) technologies have been identified as a viable alternative to improve access to care in underserved communities. Successful RPM platforms are designed and implemented for seamless integration into healthcare providers' work to increase adoption and availability for offering remote care. A quantitative survey was designed and administered to elicit perspectives from a wide range of stakeholders, including healthcare providers and healthcare administrators, about barriers and facilitators in the adoption and integration of RPM into clinical workflows in underserved areas. Ease of adoption, workflow disruption, changes in the patient-physician relationship, and costs and financial benefits are identified as relevant factors that influence the widespread use of RPM by healthcare providers; significant communication and other implementation preferences also emerged. Further research is needed to identify methods to address such concerns and use information collected in this study to develop protocols for RPM integration into clinical workflow.

**Abbreviations**

CVD: cardiovascular disease

EHR: electronic health record

HIPAA: Health Insurance Portability and Accountability Act

HIT: health information technology

KW: Kruskal-Wallis test

RPM: remote patient monitoring

Chronic medical conditions are the leading cause of death and disability in the United States (U.S.).[1] Cardiovascular diseases (CVD) and diabetes are two of the most prevalent chronic conditions. Recent estimates indicate that 1 in every 4 deaths in the U.S. is related to a CVD.[2] Additionally, approximately 11.3% of the U.S. population is affected by diabetes.[3] The economic impact of chronic disease management is significant for patients as well as the U.S. healthcare system. Approximately 90% of the annual healthcare expenses are directed towards managing chronic conditions,[4] among which CVD and diabetes account for the highest proportion of healthcare expenses.[5] In addition to monetary costs, studies have shown significant social burden, especially among family and loved ones of those affected.[6]

Chronic disease management is particularly challenging for patients in underserved communities. Studies have shown significant disparity in the prevalence of chronic diseases (e.g., CVD risks) between people residing in rural areas compared with those in urban areas.[7] Telehealth has been identified as a potential solution to provide high-quality, affordable, and timely healthcare access to patients in underserved communities.[8] For our purposes, we define telehealth broadly as a type of health information technology (HIT) that relies on telecommunication technology to deliver clinical health information and interact with providers. Telehealth systems can function in real-time (e.g., live videoconferencing), asynchronously (also known as store-and-forward), remotely (by using remote patient monitoring [RPM]), or via mobile applications.[9] Implemented telehealth platforms have shown promise in delivering healthcare to rural and remote underserved areas[8] by enhancing patient education and providing improved information access and transfer.[10,11] Despite increased adoption and demonstrated efficacy of telehealth in the last decade, underserved populations have shown lower adoption rates relative to the urban population.[12] Such low adoption rates could be due to patients' lack of access to physicians who provide telehealth services; therefore, new initiatives are needed to increase healthcare providers' adoption and use of telehealth systems.[12]

Previous research has looked at integration of telehealth systems into physicians' work. For example, a qualitative study[13] regarding the use of RPMs, one of the most common telehealth modalities to monitor patients' blood glucose and blood pressure, identified significant barriers to the successful adoption and implementation. While this and other studies have investigated the integration of RPM technologies into clinical workflow,[13-18] there is a general gap in understanding the efficacy of RPMs for medically underserved populations which lack access to primary care services and face economic, cultural, or linguistics barriers to health care.[19] Technical, organizational, social, and legal factors in underserved communities impose unique challenges (e.g., costs, data security, technical support, and patients' and providers' willingness to adopt the RPM systems).[20] Additionally, most studies focused on collecting perspectives at the operational end of a healthcare system (e.g. physicians and nurses). Few studies have elicited the perspectives of stakeholders at healthcare administrative and managerial roles to develop a holistic understanding .[21] To address these gaps, the aim of this study is to elicit the perspectives of a wide range of healthcare administrators such as executives, managers and leaders in healthcare systems as well as physicians regarding barriers and opportunities related to RPMs in an underserved population. Insights from these groups would provide valuable information to be considered in the design and development of future RPM systems for the underserved.

**Methods**

**Data collection.** A survey was developed to elicit perspectives on enablers and barriers for the adoption and integration of RPM technology in underserved communities. While the survey focused on a wide range of stakeholders including government entities, insurance companies, and relevant industries, this paper documents the responses from healthcare providers and administrators. The survey instrument included quantitative questions on topics including attitudes, beliefs, and perspectives about patients, technology acceptance, and public policy implications of adopting RPM technology. RPM was defined in the instructions as the use of digital technologies to collect medical and other forms of health data from individuals in one location and electronically transmit that information securely to healthcare providers in a different location for assessment and recommendations. The survey questions focused on five themes: (1) knowledge about the use of RPM to manage chronic diseases with special emphasis on CVD and diabetes, (2) ease of adoption and workflow disruption, (3) relationship between patients and physicians, (4) costs and financial benefits, and (5) procedures for the integration of RPM into clinical workflow, emphasizing the application in underserved populations. The data collection process was assisted by Qualtrics®, a leading marketing research firm that employs multiple methods to identify panels of potential survey respondents. The participants recruited through Qualtrics had to work in the U.S. at a managerial level in healthcare facility or as a healthcare provider (i.e., physicians). Qualtrics employs similar recruitment methods in all its panels. Individuals who qualified for the survey based on self-reported data were notified via email and invited to participate in the survey for an incentive given on a point system. The recruitment email included information including the topic of the survey, its duration (30 minutes), and a link to follow if recipients would like to participate. In addition to Qualtrics recruitment, the team was assisted by personnel of the [censored for blind review] in the identification of other stakeholders and dissemination of the survey instrument using telephone interviews. Participants in the phone interviews were paid $50 for their participation. This research was reviewed and approved by the [censored for blind review] Institutional Review Board (IRB Protocol #: IRB2017-0784D).

**Data analysis.** Mean and standard deviation of the healthcare providers' responses to scalar questions, and percentages for multiple choice, were tabulated. A bivariate analysis, using the Spearman's rank-order correlation coefficients as measure, was conducted in the cases where determining the empirical relationship between two variables was of interest. A non-parametric Kruskal-Wallis (KW) test was used to confirm significant difference in the healthcare providers' opinions, since the data did not meet the normality and homoscedasticity condition. A Games-Howell post-hoc test was performed as a follow up to identify where the significant differences resided. All analyses were performed in R studio.[22]

**Results**

Overall, 267 respondents completed the survey. The analysts reviewed each entry for completeness, thoroughness of response, and data quality. A total of 69 records were removed for partially complete or incomplete responses and unreliable responses due to unrealistic completion times. A final sample of 198 responses remained, representing healthcare providers, insurance companies, and government agencies. The scope of the current paper is limited to the analysis of healthcare providers' (n=63) perspectives; additional qualitative questions and the results from insurance and government representatives is forthcoming. The final sample of 63 responses from healthcare providers and administrators was included for analysis. Table 1 presents key demographics for the respondents included for analysis.

[Insert Table 1]

**Survey – healthcare provider responses.** Table 2 summarizes the questions of interest and basic statistics for the healthcare providers' responses. Further explanation about the results and the additional bivariate and statistical significance tests conducted are below.

[Insert Table 2]

*Knowledge about chronic health conditions management and the use of RPM*. Healthcare providers were asked to rate their knowledge about diabetes management and heart disease management, in addition to being asked to rate their knowledge about the management of chronic health conditions in general (Table 2, Question 1). Healthcare providers' responses showed a moderate level of knowledge about diabetes (M=6.73, SD=2.19), knowledge about heart disease (M=6.47, SD=2.19), and knowledge about chronic diseases in general (M=6.80, SD=2.42).

Healthcare providers were also asked to rate their level of knowledge about RPM technologies in the context of diabetes management and heart disease management (Table 2-Question 1). These responses were consistent across both disease types, with no major difference between their perceived knowledge levels about the use of RPM for diabetes management (M=4.98, SD=3.13) and the use of RPM for heart disease management (M=4.98, SD=3.05). Responses to the level of knowledge about RPM of diabetes showed a relatively uniform distribution, and a slight crest between '7' and '8'. The distribution of responses was similar for perceived level of knowledge about RPM of CVD.

*Ease of adoption and workflow disruption.* To investigate the concepts of ease of adoption and disruption to clinical workflow in the context of the use of RPM for underserved communities, we asked the surveyed healthcare providers to indicate their perspectives about these topics (Table 2, Question 2). Results suggest that healthcare providers are modestly positive about the ease of adoption of RPM systems (M=6.12, SD=2.14) and somewhat neutral whether RPMs will result in disruption to clinical workflows (M=5.37, SD=2.22). A bivariate analysis was performed to investigate the relationship between levels of knowledge about RPM and perspectives regarding ease of adoption and disruption to clinical workflow further (Figure 1). The analysis showed that perspectives about disruption to clinical workflow, and knowledge levels about the use of RPM for diabetes and heart disease management had a weak positive correlation (Spearman's rank-order correlation coefficients of 0.06 and 0.05, respectively). In contrast, perceived ease of RPM adoption and levels of knowledge in chronic disease management (diabetes and heart disease) had a moderate positive correlation (Spearman's rank-order correlation indexes of 0.52 and 0.55, respectively).

[Insert Figure 1]

*Relationship between patients and physicians.* To understand participants' perspectives on the effect of RPM on the physician-patient relationship in underserved communities, healthcare providers were asked to rate how helpful (or unhelpful) they perceived RPM technologies would be to enhance the patient-physician case management relationship (Table 2, Question 3). Healthcare providers' responses suggest that they perceived RPM as beneficial in enhancing the patient-physician relationship in underserved communities (M=6.80, SD=1.71). Regarding healthcare providers' level of trust in patients' adherence to RPM-based medical recommendations (Table 2, Question 2), results reveal that healthcare providers seemed to be neutral in their response (M=5.73, SD=2.09).

*Cost and financial benefits.* Healthcare providers were asked to provide their perspective on the cost of RPM technology (Table 2, Question 4) and the potential financial benefits they could receive from adopting an RPM system (Table 2, Question 2). Our results indicate that healthcare providers seemed to be in agreement that using RPM technology would be costly for providers (M=6.53, SD=1.68). Healthcare providers' responses were less optimistic regarding the financial benefits they would receive from adopting an RPM system (M=5.15, SD=2.15; see Figure 2).

[Insert Figure 2]

The surveyed healthcare providers were asked to provide their opinion about the suitability of five methods of payment for reimbursing physician who use RPM to manage their patients' care (Table 2, Question 5). Survey responses for this question show that while healthcare providers seem to be in favor of private medical insurance (M=6.98, SD=1.91), Medicare (M=6.84, SD=2.14), and Medicaid (M=6.41, SD=2.52) as suitable methods for reimbursement, they are less optimistic regarding contract with RPM device manufacturer (M=5.85, SD=2.06) and not in favor of patients' out-of-pocket expense (M=4.85, SD=2.50). Results from the KW test showed a significant difference between the means (P<.001) of the reimbursement alternatives presented to the healthcare providers. The Games-Howell post-hoc test (Table 3) showed that patients' out-of-pocket expense was rated significantly lower by the healthcare providers compared with private medical insurance (P<.001) and Medicare (P=.003). Overall, healthcare providers seem to primarily support private medical insurance as a method of reimbursement, in addition to public payer alternatives such as Medicare and Medicaid (Figure 3). The post-hoc test revealed no significant difference between these three methods of reimbursement.

[Insert Table 3]

[Insert Figure 3]

*Time allocation for RPM data review*. To understand healthcare providers' perceptions and expectations for adequate time allocation for RPM activities, the surveyed healthcare providers were asked to indicate the percentage of a healthcare provider's working time each day that should be spent reviewing RPM data notifications (Table 2, Question 6). Response values ranged from 3% to 90% which suggests large variability in perceptions. Overall, the most commonly expected percentage of working time was below 30%.

*RPM Data communication and representation*. To gain insights about healthcare providers' preferences regarding the communication method to receive the summarized patient health data and the form of presentation, the surveyed healthcare providers were asked to rate eight ways for a healthcare provider to receive patients' remote health information (Table 2, Question 7) and five ways for a healthcare provider to be presented with patients' remote readings of blood glucose levels (Table 2, Question 8). Out of the eight alternatives presented for ways to receive RPM data, the healthcare providers showed a preference for email (M=3.75, SD=1.31) and telehealth computer systems (M=3.59, SD=1.29) as ways to receive patients' RPM data (Figure 4). Results from the KW test showed significant difference between the means (P<.001) of the healthcare providers responses. The corresponding Games-Howell post-hoc test (Table 4) confirmed a significant mean difference (P<.05) of the two preferred alternatives (email and telehealth computer systems) in comparison with other alternatives such as phone calls, fax, and listening ports. However, no such significant difference resulted when both methods were compared with the use of mobile apps as a way for healthcare providers to receive RPM data.

[Insert Figure 4]

[Insert Table 4]

Regarding the ways for a healthcare provider to be presented with patients' remote readings of blood glucose levels, the non-parametric KW test showed a significant difference between the means (P<.001) of the alternatives presented to the healthcare providers. The results from the healthcare responses suggest a preference for receiving the RPM data in table/chart format (M=3.77, SD=1.14) or in picture/graph (M=3.60, SD=1.23; see Figure 5); however, this is statistically inconclusive since the Games-Howell post-hoc test (Table 5) revealed no significant mean differences with the other alternatives such as text message format and in-person through office assistant. On the other hand, the post-hoc test showed significant mean differences between auditory alternatives such as voice notes in comparison with the other alternatives such as picture/graph, table/chart, and in-person through office assistant.

[Insert Figure 5]

[Insert Table 5]

*When the RPM data should be received*. The healthcare providers were asked to indicate their preference regarding the best time of the day to receive communication from an RPM system about overall patient health status (Table 2, Question 9). The participants were presented with five alternatives. Results show that 40% of the participants favored receiving RPM data at the beginning of the work day.

**Discussion**

Our investigation provided insights into barriers for healthcare providers' adoption of RPM technologies and significant aspects of the technology integration into clinical workflow, with a specific focus in the context of RPM for underserved communities. The survey responses mainly represent the perspectives of well-educated and experienced executives and managers from healthcare settings such as hospital and clinics, an aspect that sets our findings apart since research in this area has focused on collecting perspectives at the operational end of a healthcare system (e.g. physicians and nurses).

**Ease of adoption and workflow disruption.** Previous research has identified concerns regarding ease of adoption and workflow disruption as relevant barriers for healthcare providers RPM adoption, therefore affecting the technology integration into clinical workflow.[13,23] In the context of underserved communities, our findings suggest that healthcare providers are modestly positive about ease of adoption. Furthermore, findings from our bivariate analysis suggest that healthcare providers who perceived themselves as more knowledgeable about RPM management seemed to have a more positive and optimistic outlook about the ease of adoption of RPM in clinical settings. These results highlight the need to investigate whether enhancing healthcare providers' knowledge through education about the benefits of adopting an RPM system may improve their perception about ease of adoption. Additionally, there is potential to investigate early exposure of healthcare providers to RPM encounters as a mechanism to increase telehealth adoption. Early exposure to RPM encounters, in addition to well-established technology integration protocols, also have the potential to decrease healthcare providers concerns regarding workflow disruption. Our findings suggest that healthcare providers are somewhat neutral whether RPMs will result in disruption to clinical workflows.

**Relationship between patients and physicians.** The potential negative impact of RPM in the patient-physician relationship is another significant barrier that has been identified in the literature for healthcare providers adoption and integration of RPM technologies. Studies have recognized that the adoption of telehealth technologies such as RPM can have a detrimental effect on doctor-patient relationship such as "depersonalization" of the treatment compared with conventional face-to-face sessions.[11] However, in the context of underserved communities, telehealth presents the potential of enhancing this valuable relationship, since it provides the opportunity to reach patients who have limited to no access to quality healthcare delivery.[24] Findings from the data analysis support that healthcare providers perceive RPM as beneficial in enhancing the patient-physician relationship in underserved communities. This optimism is relevant since the quality of the patient-physician relationship can influence patient compliance with medical advice.[25] On the other hand, our findings suggest that healthcare providers seem to be neutral regarding the utility of RPM in ensuring patient treatment adherence. Historically, patients' adherence to physicians' recommendations has been a challenge that healthcare systems have encountered even in traditional face-to-face encounters,[26] which may explain healthcare providers' expectation to encounter similar issues with remote patient care. However, as the adoption and implementation of RPM systems increases, work is warranted to investigate if RPM shifts providers' perception of patients' compliance behavior and patients' actual behavior.

**Cost and financial benefits**. Despite the existence of success stories where the implementation of RPM services has resulted in cost savings and positive return on investment,[27] evidence supporting such cost-effectiveness is limited.[28] Consistent with findings from the literature,[29] our respondents agreed that using RPM technology would be costly for providers. Additionally, results suggest that uncertainties about the financial incentives and reimbursement for physicians may affect healthcare providers' perception about benefits outweighing the associated costs. This is in line with previous research where healthcare providers have expressed their concerns about the prohibitive upfront costs and time for return on initial investments[13] as well as coverage and reimbursement being major obstacles that present greater challenges for physicians' adoption of telehealth systems.[26,30] Healthcare providers' preference to primarily support private medical insurance, in addition to public payer alternatives, as primary methods of reimbursement reflects the actual panorama on telehealth services in the U.S. Many states have passed laws that requires private payers to provide coverage and reimbursement for telehealth services. On the public payer side, more inconsistencies exist due to the restrictive coverage and reimbursement policies of Medicare and Medicaid, regarding statutory barriers, limitations in the types of practitioners that can use telehealth services, and restrictions of reimbursement for only real-time interactive encounters.[26,31] However, telehealth laws and reimbursement policies vary by state, and as of spring 2022, 30 state Medicaid programs provide some form coverage and reimbursement for RPM services.[32]

**The Importance of Workflow Implementation Preferences.** While telehealth integration has been studied in time/motion studies in terms of activity flow and sequence,[15-17] to the best of our knowledge there is a lack of efforts in investigating healthcare providers' preferences that may be relevant for the development of workflow procedures such as regarding frequency and timing of RPM data injects, communication types, and representation of RPM data. Understanding these factors is essential to mitigate data overload and interruptions in clinical workflow, and to assure means and ways to present RPM data fits healthcare providers' mental models.

*Time allocation for RPM data review.* Results from time studies in the context of RPM implementation have revealed that processing remote transmissions is faster when compared with in-person evaluations.[16] Additionally, studies have identified that an RPM approach (specifically for remote follow-up of cardiac implantable devices) requires less workforce, representing a gain in time and resources in comparison with in-hospital follow-ups.[17] However, despite identifying these benefits in comparison to in-person visits, studies have not evaluated whether the time spent by healthcare providers processing and evaluating RPM data is consistent with their expectations. In previous studies physicians have described the task of reviewing RPM data as being as seamless as reading a nurse's note,[33] suggesting that physicians would not expect to spend much of their working time reviewing RPM data. Our findings suggest that on average, healthcare providers expect to spend 30% or less of their working time reviewing RPM data. However, it is important to note that our findings mostly represent the perspectives of executives and managers at leadership levels in a healthcare system hierarchy. Future studies must elicit physicians' perspectives about this issue to understand expectations about time allocation from an operational point of view.

*RPM Data communication and representation.* In previous research, healthcare providers have expressed a strong preference for viewing summarized telemonitoring data instead of receiving raw data.[13] However, there is no evidence regarding the healthcare providers' preferences regarding the communication method to receive the summarized patient health data and the form of presentation. Findings from our data analysis suggest healthcare providers prefer email and telehealth computer systems as ways to receive patients' RPM data. Emails are one of the main ways of communication in the modern corporate world, however, state laws rarely accept emails as a main form of healthcare delivery for reimbursement unless it is used in conjunction with some other type of system, such as a telehealth system.[32] Using this mean of communication for sharing patient data would require data warehousing and security standards that are Health Insurance Portability and Accountability Act (HIPAA) compliant. Healthcare providers' preference for the use of a telehealth computer system suggests their desire to have a central repository where they can access remote patient data. However, the platform should be interoperable with other HIT used in the clinics, such as EHRs, to facilitate the evaluation and charting activities.[13] Our data analysis showed inconclusive results regarding healthcare providers preferred alternative for ways to be presented with patients' RPM data; however, auditory alternatives such as voice notes were significantly least preferred possibly due to their potential for creating disruption in the workplace.

*When the RPM data should be received.* Establishing when the data should be presented to the healthcare provider is essential to avoid interruptions and data overload at inconvenient times. Our findings suggest that healthcare providers prefer to receive patients' data at the beginning of the workday or at a pre-scheduled time during the workday. Receiving patients' data at the beginning of the day could be preferred because it provides opportunity for the provider to take timely action (i.e. follow-up or to schedule an in-person visit) when presented with alarming or abnormal values in patients' health data. Preference to receive communication at a pre-scheduled time suggests that healthcare providers desire flexibility in planning their workflow. For example, they may desire to customize when to receive the data based on attributes such as patient condition and severity of the condition.

**Limitations.** Results from the survey mainly represent the viewpoint of executives and managers of healthcare settings, which correspond to 59% of the survey respondents. Although the results from the survey provide significant insights about healthcare providers' perspectives on the impact of the integration of RPMs into their work in underserved communities, more data must be collected to capture physicians' perspectives in depth. The low n-count of physicians in the sample prevents running desirable cross-group or sub-group analyses at this time. Also, further analysis showed that the healthcare providers surveyed work in healthcare settings primarily located in urban areas. Future work is needed to capture the perspectives of healthcare providers working in underserved rural areas.

**Conclusion**

An exploratory analysis of healthcare provider responses provided insights about their perspectives regarding barriers and facilitators for RPM to transform healthcare access in underserved communities, and their viewpoints regarding aspects to consider when developing protocols for the integration of RPM into clinical workflow. While healthcare providers showed optimism about the impact of RPMs on doctor-patient relationships and the ease of adoption of RPM systems in underserved communities, they were apprehensive about the consequent disruption to clinical workflow. Cost of adoption was also perceived as a major factor contributing to healthcare provider apprehension towards RPM integration. Further research efforts must be directed towards identifying methods to address such concerns and acquiring a better understanding of what are the implications for policy and implementation of RPM interventions.

Overcoming the aforementioned barriers to achieve a successful adoption and integration of RPM systems in clinical workflow implies the development of efficient workflow procedures for large-scale implementation. The integration of new health information technologies such as RPM will have an impact on traditional clinical workflow, as well as the roles and functions of physicians and clinical staff members. Such disruptions can cause an impact in the health provider productivity and time efficiency, quality of care, and patient safety as clinics move to larger-scape implementation of telehealth systems. Results from the survey provided insights about healthcare providers' perspectives regarding aspects of timing, communication, and representation of RPM data. Further research endeavors should be directed towards using this information to develop protocols for RPM integration in clinical workflow, in addition to identifying other aspects that must be considered.

**Acknowledgments**

[censored for blind review]

**Conflicts of Interest**

The author(s) disclosed receipt of the following financial support for the research, authorship, and/or publication of this article: This work is sponsored by the National Science Foundation [censored for blind review].

Table 1. Key demographics of survey population - healthcare providers

| Characteristic | n (%) |
|---|---|
| **Gender** | |
| Female | 27 (42.9) |
| Male | 36 (57.1) |
| **Age** | |
| 18-34 years | 4 (6.3) |
| 35-44 years | 12 (19.0) |
| 45-54 years | 14 (22.2) |
| 55-64 years | 21 (33.3) |
| 65-74 years | 8 (12.7) |
| 75 or more years | 4 (6.3) |
| **Race/ethnicity** | |
| White | 53 (84.1) |
| Hispanic or Latino | 3 (4.8) |
| Black or African American | 1 (1.6) |
| Asian | 5 (7.9) |
| Prefer not to answer | 1 (1.6) |
| **Education** | |
| Associate degree | 8 (12.7) |
| Bachelor's degree | 15 (23.8) |
| High school diploma or GED | 2 (3.2) |
| Master's degree | 9 (14.3) |
| Professional, medical, or doctorate degree | 29 (46.0) |
| **Job title** | |
| Executive/C-Suite | 12 (19.0) |
| Manager (less than 3 years of experience) | 5 (7.9) |
| Manager (more than 3 years of experience) | 20 (31.7) |
| Contributor with experience in healthcare | 10 (15.9) |
| Leader (looks after a region or business area) | 5 (7.9) |
| Physician | 5 (7.9) |
| Other | 6 (9.5) |

**Table 2.** Survey key questions analyzed

| Question | Mean | Standard Deviation |
|---|---|---|
| **1. How much do you think you know about each of the following?** Use the SLIDING SCALE of 0 to 10 for your response, where 0 is Know Nothing About It and 10 is Know A Great Deal About It. | | |
| • Chronic health conditions management | 6.80 | 2.42 |
| • Diabetes management | 6.73 | 2.19 |
| • Heart disease management | 6.47 | 2.19 |
| • Use of RPM for diabetes management | 4.88 | 3.13 |
| • Use of RPM for heart disease management | 4.98 | 3.05 |
| **2. Indicate your level of disagreement or agreement with the following statements regarding RPM systems.** Use the SLIDING SCALE of 0 to 10 for your response, where 0 is Completely Disagree and 10 is Completely Agree. | | |
| • A healthcare provider could easily adopt an RPM system. | 6.12 | 2.14 |
| • Adopting an RPM system would represent a major disruption to a healthcare provider's daily work. | 5.37 | 2.22 |
| • A healthcare provider will trust that patients will follow medical recommendations communicated through the RPM system. | 5.73 | 2.09 |
| • A healthcare provider will benefit financially from the implementation of an RPM system. | 5.15 | 2.15 |
| **3. How unhelpful or helpful do you think RPM would be with each of the following patient health benefits?** Use the SLIDING SCALE of 0 to 10 for your response, where 0 is Extremely Unhelpful and 10 is Extremely Helpful. | | |
| • Enhance doctor-patient case management relationship. | 6.80 | 1.71 |
| **4. Indicate how much you disagree or agree with the following statements.** Use the SLIDING SCALE of 0 to 10 for your response, where 0 is Completely Disagree and 10 is Completely Agree. | | |
| • RPM technology is costly for providers. | 6.53 | 1.68 |
| **5. How unsuitable or suitable are the following methods of payment for reimbursing physicians who use RPM to manage their patients' care?** Use the SLIDING SCALE of 0 to 10 for your response, where 0 is Extremely Unsuitable and 10 is Extremely Suitable. | | |
| • Private medical insurance | 6.98 | 1.91 |
| • Medicare | 6.84 | 2.14 |
| • Medicaid | 6.41 | 2.52 |
| • Patient's out-of-pocket expense | 4.85 | 2.5 |
| • Contract with RPM device manufacturer to provide service | 5.85 | 2.06 |
| **6. What percentage of a health practitioner's working time each day should be spent reviewing RPM data notifications?** Use the SLIDING SCALE of 0 to 100 percent for your response. | 36.96 | 25.04 |

7. **How preferable do you think the following ways are for a physician to receive patients' remote health information?**

Please rate each of the following on a 5-point scale ranging from least preferred to most preferred.

|  |  |  |
|---|---|---|
| • Email | 3.75 | 1.31 |
| • Phone call | 2.44 | 1.54 |
| • Smartphone texting | 2.95 | 1.31 |
| • Fax | 2.58 | 1.29 |
| • Tablet PC | 2.98 | 1.28 |
| • Telehealth computer system | 3.59 | 1.29 |
| • Mobile app | 2.98 | 1.25 |
| • Listening ports | 2.57 | 1.18 |

8. **How preferable do you think the following ways are for a physician to be presented with patients' remote readings of blood glucose levels?**

Please rate each of the following on a 5-point scale ranging from least preferred to most preferred.

|  |  |  |
|---|---|---|
| • Text message format | 3.23 | 1.45 |
| • Picture/Graph | 3.60 | 1.23 |
| • Table/Chart | 3.77 | 1.14 |
| • Voice Notes | 2.66 | 1.16 |
| • In-person though office assistant | 3.31 | 1.17 |

9. **What time of day is best for a healthcare provider to receive communication from an RPM system for all patients about overall patient health status?** — **Percentage (%)**

| | |
|---|---|
| • At the beginning of the work day | 40 |
| • At a pre-scheduled time during the work day | 30 |
| • On demand access | 13 |
| • At the end of the work day | 9 |
| • Don't know | 9 |

**Table 3**. Games-Howell post-hoc test for methods of payment for reimbursing physician who use RPM to manage their patients' care

| Groups | *P* value |
| --- | --- |
| Private medical insurance : Medicare | .993 |
| Private medical insurance : Medicaid | .594 |
| Private medical insurance : Patient's out-of-pocket | <.001* |
| Private medical insurance : RPM device manufacturer | .078 |
| Medicare : Medicaid | .843 |
| Medicare : Patient's out-of-pocket | .003* |
| Medicare : RPM device manufacturer | .255 |
| Medicaid : Patient's out-of-pocket | .102 |
| Medicaid : RPM device manufacturer | .928 |
| Patient's out-of-pocket : RPM device manufacturer | .302 |

*Statistically significant (p-value < .05)

**Table 4**. Games-Howell post-hoc test for ways for a healthcare provider to receive patients' remote health information

| Groups | *P* value |
| --- | --- |
| Email : Phone call | .005* |
| Email : Smart phone texting | .171 |
| Email : Fax | <.001* |
| Email : Tablet PC | .166 |
| Email : Telehealth computer system | .998 |
| Email : Mobile app | .119 |
| Email : Listening Ports | .001* |
| Phone call : Smart phone texting | .805 |
| Phone call : Fax | 1 |
| Phone call : Tablet PC | .801 |
| Phone call : Telehealth computer system | .031* |
| Phone call : Mobile app | .820 |
| Phone call : Listening Ports | 1 |
| Smart phone texting : Fax | .522 |
| Smart phone texting : Tablet PC | 1 |
| Smart phone texting : Telehealth computer system | .529 |
| Smart phone texting : Mobile app | 1 |
| Smart phone texting : Listening Ports | .667 |
| Fax : Tablet PC | .514 |
| Fax : Telehealth computer system | .004* |
| Fax : Mobile app | .530 |
| Fax : Listening Ports | 1 |
| Tablet PC : Telehealth computer system | .522 |
| Tablet PC : Mobile app | 1 |
| Tablet PC : Listening Ports | .660 |
| Telehealth computer system : Mobile app | .436 |
| Telehealth computer system : Listening Ports | .007* |
| Mobile app : Listening Ports | .679 |

*Statistically significant (p-value < 0.05)

**Table 5**. Games-Howell post-hoc test for ways for a healthcare provider to be presented with patients' remote readings of blood glucose levels

| Groups | P value |
| --- | --- |
| Text message format : Picture/graph | .568 |
| Text message format : Table/chart | .201 |
| Text message format : Voice notes | .115 |
| Text message format : In-person through office assistant | .994 |
| Picture/graph : Table/chart | .959 |
| Picture/graph : Voice notes | <.001* |
| Picture/graph : In-person through office assistant | .745 |
| Table/chart : Voice notes | <.001* |
| Table/chart : In-person through office assistant | .294 |
| Voice notes : In-person through office assistant | .016* |

*Statistically significant (p-value < 0.05)

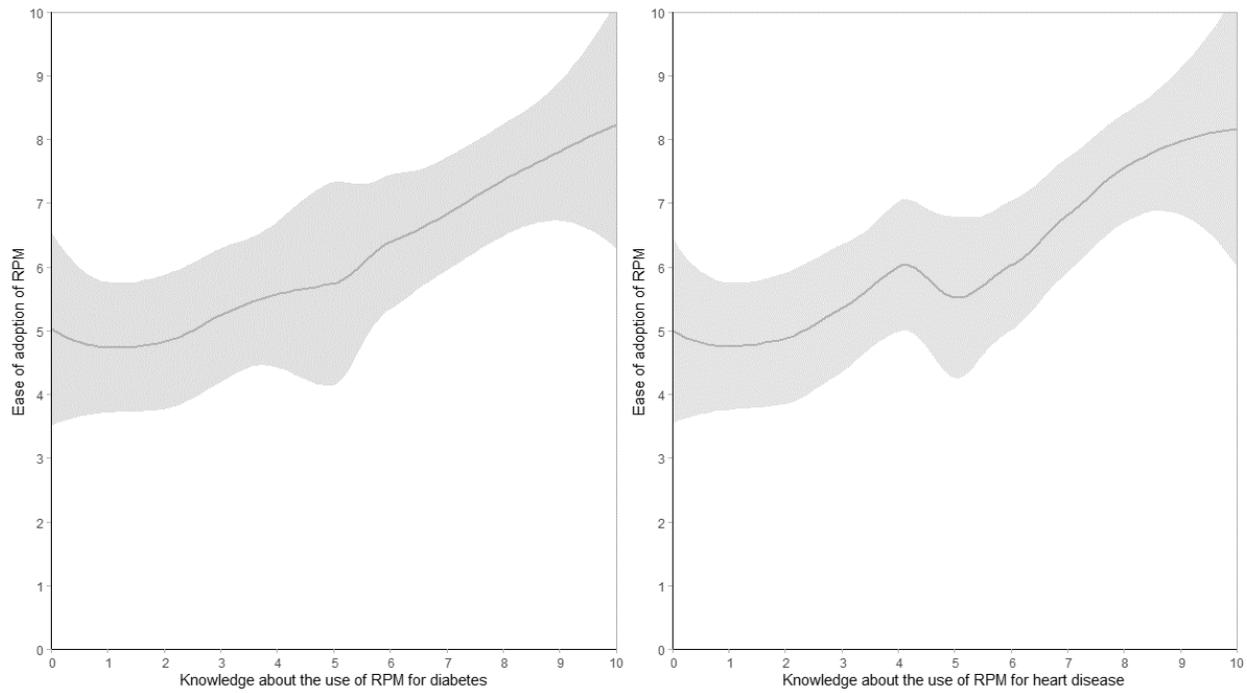

**Figure 1**. Bivariate Analysis for Ease of adoption and ease of adoption of RPMs vs. knowledge about RPM use for diabetes (left) and heart disease (right)

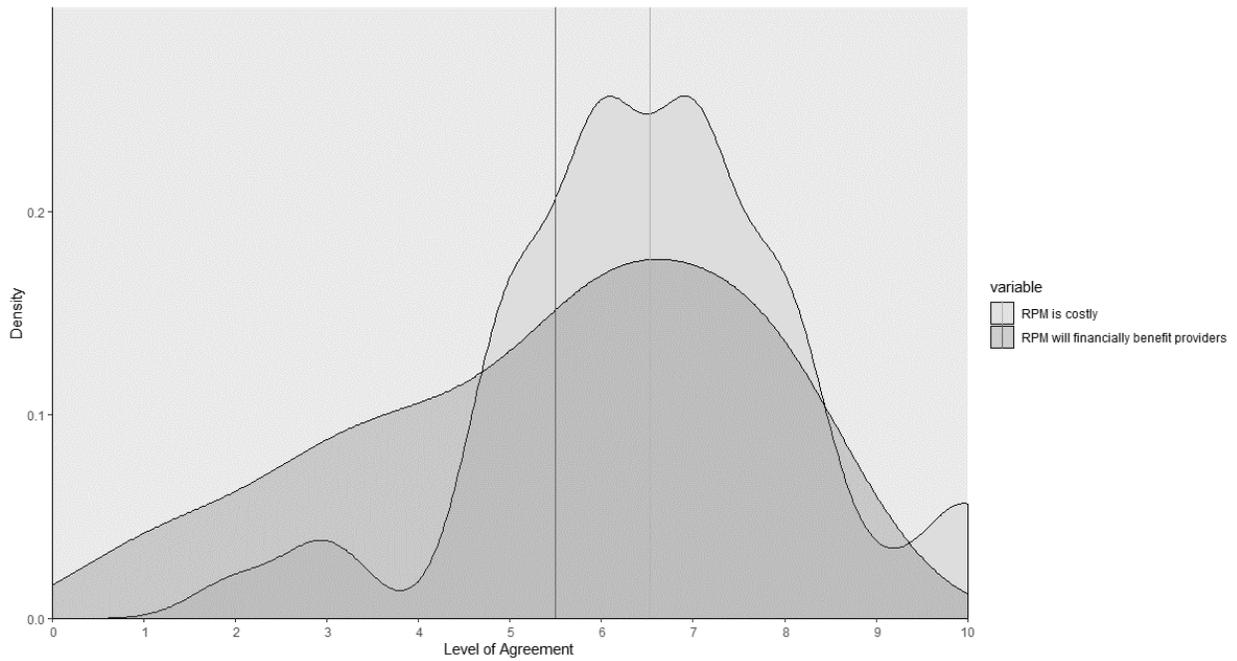

**Figure 2**. Distribution of responses to perceptions regarding the cost and financial benefits of RPM for healthcare providers

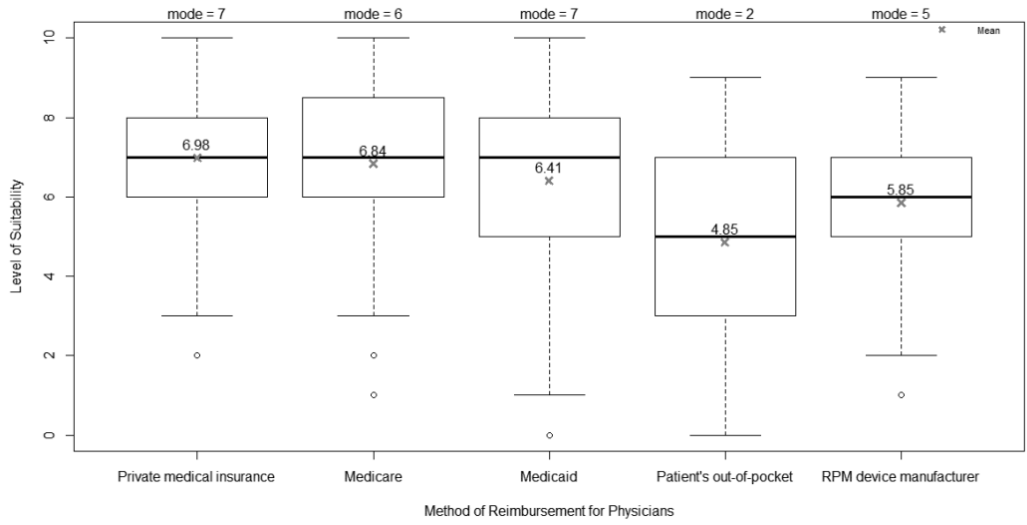

**Figure 3**. Comparison of perceived level of suitability of different methods for reimbursement

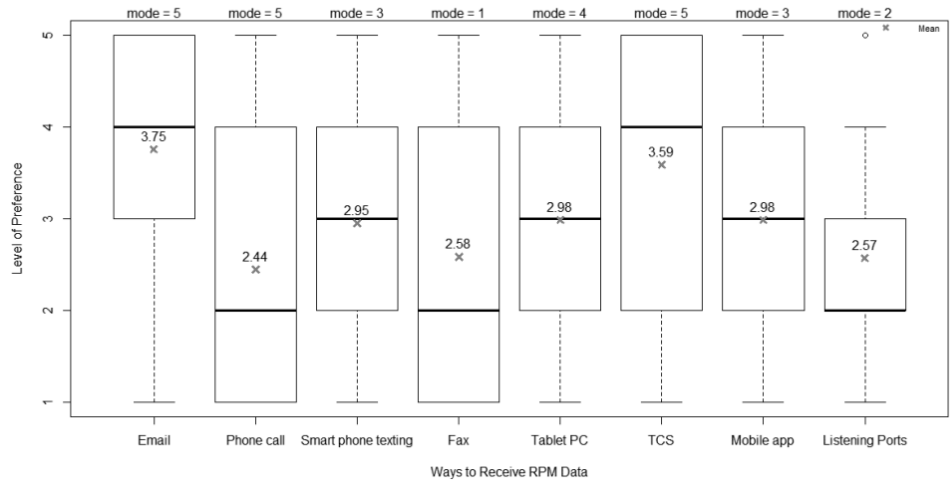

**Figure 4**. Methods to receive RPM data

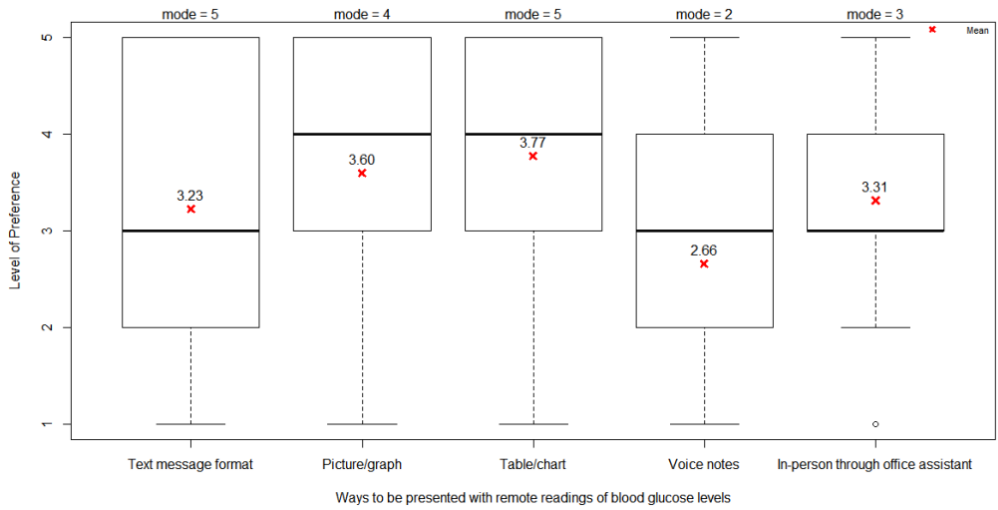

**Figure 5**. Ways to be presented with remote readings of blood glucose levels